\documentclass{jpp}
\usepackage[T1]{fontenc}
\usepackage[utf8]{inputenc}
\usepackage{amsmath,amssymb}
\usepackage{graphicx}
\usepackage{url}
\usepackage{color}
\usepackage[hidelinks]{hyperref}
\hypersetup{pdftitle={Analytic toroidal 3D MHD equilibria and steady Euler flows with invariant surfaces},
  pdfauthor={Matt Landreman},
  pdfsubject={Analytic toroidal 3D MHD equilibria and steady Euler flows with invariant surfaces},
  pdfkeywords={MHD equilibrium, stellarators, flux surfaces, Grad's conjecture}}

\newcommand{\B}{\mathbf{B}}
\newcommand{\rvec}{\mathbf{r}}
\newcommand{\xvec}{\mathbf{x}}
\newcommand{\eR}{\mathbf{e}_R}
\newcommand{\ephi}{\mathbf{e}_\phi}
\newcommand{\eZ}{\mathbf{e}_Z}
\newcommand{\eps}{\epsilon}
\newcommand{\thetageo}{\theta}

\shorttitle{Analytic toroidal 3D MHD equilibria \& steady Euler flows with invariant surfaces}
\shortauthor{M. Landreman}
\title{Analytic toroidal 3D MHD equilibria and steady Euler flows with invariant surfaces}
\author{Matt Landreman\aff{1}\corresp{\email{mattland@umd.edu}}}
\affiliation{\aff{1}University of Maryland, College Park, MD 20742, USA}

\begin{document}
\maketitle

\begin{abstract}
Families of explicit analytic solutions of the magnetohydrodynamic equilibrium equations are presented, equivalent to steady incompressible Euler flow.
The solutions are non-axisymmetric and possess exact nested toroidal flux surfaces.
No expansion is made in inverse aspect ratio or in the deviation from axisymmetry.
The magnetic field and scalar pressure are given explicitly in Cartesian coordinates using elementary functions.
The field, current density, and pressure are smooth over the toroidal domain.
The pressure gradient vanishes only on the magnetic axis.
One family of solutions has uniform rotational transform $\iota=2$, while another family has a sheared $\iota$ profile.
These counterexamples to Grad's conjecture are valuable for understanding the existence and regularity of 3D equilibria and for testing numerical codes.

\end{abstract}


\section{Introduction}
\label{sec:introduction}

Three-dimensional (3D) magnetohydrodynamic (MHD) equilibria are central to both the stellarator and tokamak concepts for magnetic-confinement fusion energy \citep{KruskalKulsrud}.
The existence of such equilibria is particularly important for stellarators, whose confinement relies on strongly non-axisymmetric  geometry, while tokamaks too have unavoidable and sometimes intentional deviations from axisymmetry.
The MHD equilibrium equations are also of interest due to their isomorphism with steady incompressible Euler flow \citep{Moffatt_1985, helander2014theory}, identifying the magnetic field with the fluid velocity.
The equilibrium equations are
\begin{equation}
(\nabla\times\B)\times\B=\nabla p,
\qquad
\nabla\cdot\B=0,
\label{eq:mhd}
\end{equation}
for magnetic field $\B$ and scalar pressure $p$.
The $\B$ component of the first equation gives $\B\cdot\nabla p=0$, so pressure surfaces are invariant surfaces for magnetic field line flow.
For magnetic confinement, these surfaces should form nested tori, with $p$ decreasing from the core to the edge.
The difficulty of satisfying this condition in non-symmetric geometry led \cite{grad1967} to conjecture that smooth solutions with non-constant pressure do not exist unless there is an additional symmetry such as axisymmetry: 
``...we find it very unlikely that there exist a general class of toroidal equilibria with smooth $p\ldots$''
He elaborated on the conjecture in \citep{grad1985theory}: ``other than the stated symmetric exceptions, there are no \emph{families} of solutions depending smoothly on a parameter.''
Variations of these statements and related questions are subjects of active study \citep{constantin2021flexibility,constantin2023magnetic, enciso2025, cardona2025asymmetry, drivas2025, peralta2026symmetry}.
Very recently, \cite{gomezserrano2026} falsified the conjecture by proving existence of non-axisymmetric solutions, with poloidally closed field lines everywhere and vanishing $\B$ on the magnetic axis.
Here we will present two parameterized classes of non-symmetric analytic solutions, for which $\B$ and $p$ can be expressed explicitly.


Previously, solutions of (\ref{eq:mhd}) have been found by relaxing conditions necessary for practical confinement or by introducing approximations.
For instance, expansions in small distance from the magnetic axis (high aspect ratio) led to approximate solutions by \cite{Mercier, SolovevShafranov, GB1}.
\cite{lortz1970} established the existence of finite-aspect-ratio nonsymmetric toroidal equilibria, but only for cases with mirror-reflection symmetry in which the rotational transform $\iota=0$, which is not compatible with magnetic confinement.
That work also did not provide any explicit solutions to (\ref{eq:mhd}), only a proof that a certain iteration should converge to some solution.
\cite{weitzner2020} found exact nonsymmetric closed-line vacuum fields in a topological torus instead of true toroidal geometry.
Nonsymmetric steady Euler flows with closed trajectories have been constructed in periodic cylindrical geometry by \cite{drivas2025}.
Explicit three-dimensional MHD equilibria with a straight magnetic axis were also obtained by \cite{kaiser1997}.
A different approach requires discontinuous pressure and current sheets \citep{bruno1996,enciso2025}.
Analytical descriptions based on linearizing about axisymmetry with a small 3D perturbation have also been investigated \citep{plunk2020perturbing,sorokina_ilgisonis2024}.

The purpose of this paper is to present fully non-axisymmetric, smooth, toroidal, finite-aspect-ratio, explicit analytic solutions possessing both exact nested flux surfaces and finite rotational transform.
All quantities are smooth, with the pressure gradient vanishing only on the magnetic axis.
The solutions here do not possess plane-reflection symmetry, which was used by \cite{lortz1970} but which is not compatible with stellarator confinement.
The configurations are stellarator-symmetric (rotation by $\pi$ about the $x$ axis), with two field periods (rotation by $\pi$ about the $z$ axis).
Two distinct families of solutions are given here.
The first has closed field lines, with $\iota=2$  everywhere.
The second family has a sheared rotational transform profile.
Both families are parameterized by aspect ratio and the departure from axisymmetry, with additional parameters in the second family.
Each family connects continuously to an axisymmetric limit.
The integer-$\iota$ family is presented in section \ref{sec:integer_family},
and the sheared-$\iota$ family is presented in section \ref{sec:sheared_family}.
Scripts confirming that numerical equilibria match these analytic solutions can be found in the supplemental material \citep{github}.

\section{Solutions with integer rotational transform}
\label{sec:integer_family}


\subsection{Magnetic field and pressure}
\label{sec:construction}

Choose $0<\eps<1$ and define $a=\sqrt{1+\eps}$ and $b=\sqrt{1-\eps}$.
The three-dimensional magnetic field in Cartesian coordinates is
\begin{equation}
\B=\left(
\frac{2zx-(a/b)Fy}{s},\;
\frac{2zy+(b/a)Fx}{s},\;
1-s
\right),
\label{eq:B_cartesian}
\end{equation}
with
\begin{equation}
s=\frac{x^2}{a^2}+\frac{y^2}{b^2},
\qquad
F=\sqrt{1 - (1-s)^2 - 4z^2}>0.
\label{eq:auxiliary}
\end{equation}
The initial domain considered is the region where the radicand is positive, $\mathcal U_\eps=\{\xvec:1 - (1-s)^2 - 4z^2>0\}$.
In this domain $s>0$, so the field is smooth.

\begin{figure}
\centering
\includegraphics[width=0.53\textwidth]{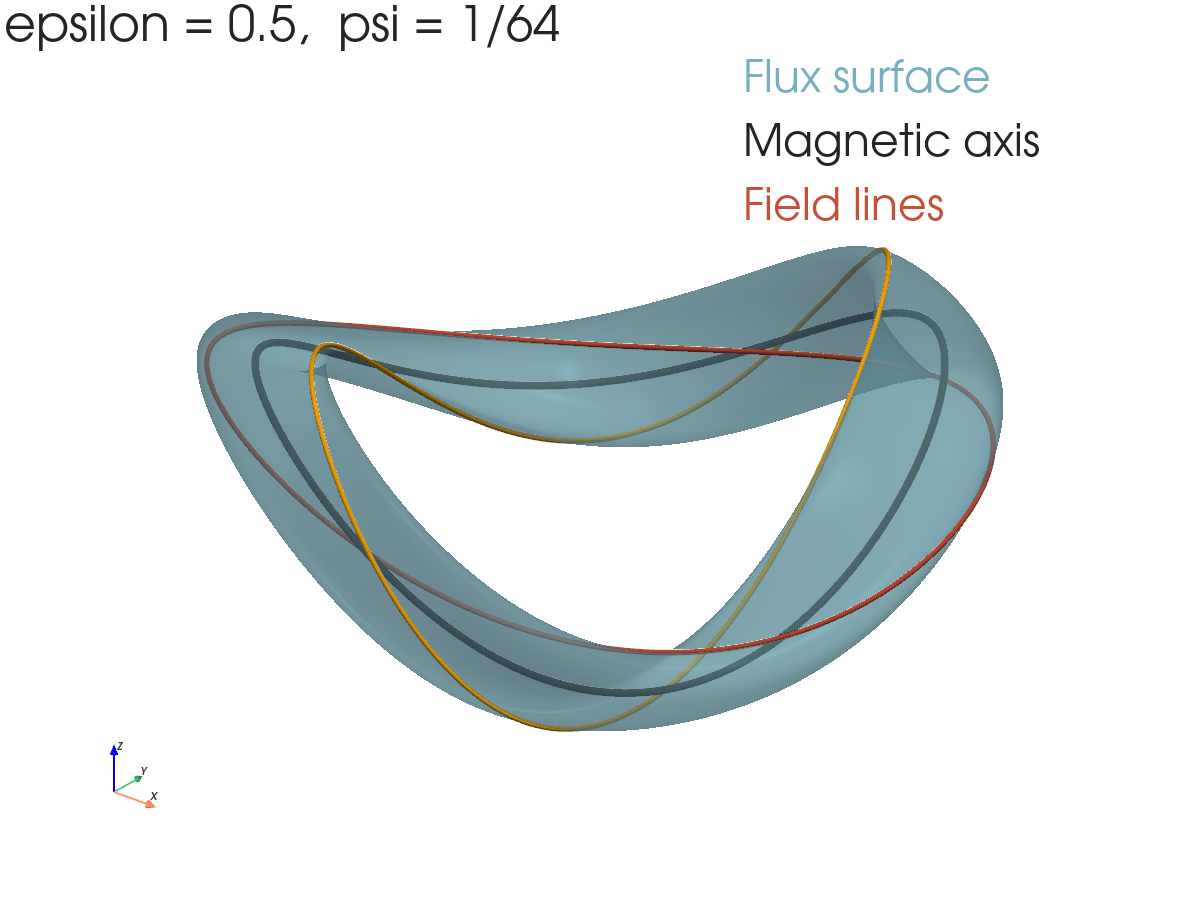}\hfill
\includegraphics[width=0.45\textwidth]{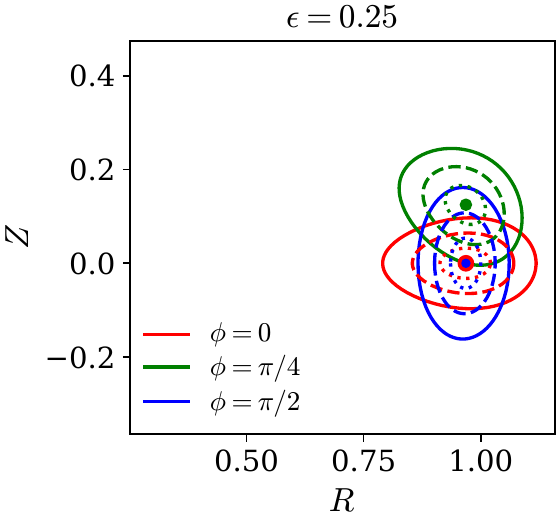}
\includegraphics[width=0.45\textwidth]{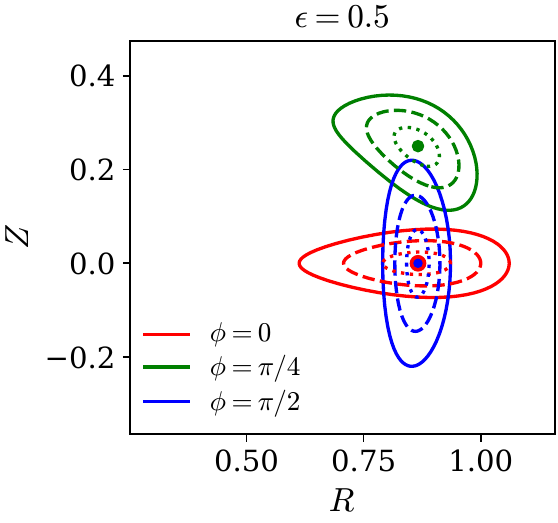}
\hfill
\includegraphics[width=0.45\textwidth]{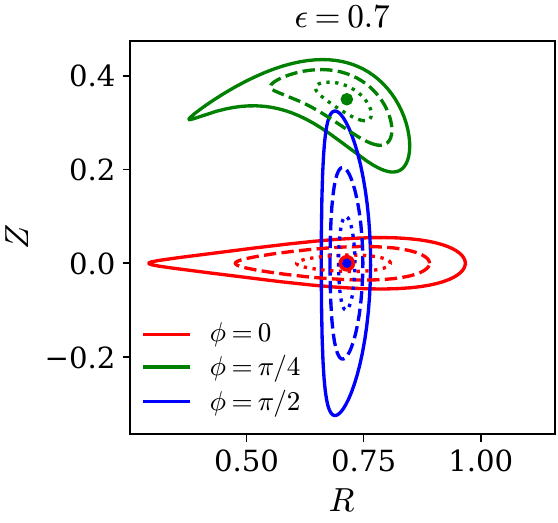}
\caption{The equilibrium family with $\iota=2$.
Top left: for $\eps=1/2$, the surface $\psi=\delta=1/64$, two field lines on this surface, and the magnetic axis.
Other panels: cross-sections of the flux surfaces for $\psi=\delta$, $4\delta/9$, and $\delta/9$.}
\label{fig:geometry}
\end{figure}

To obtain the flux surfaces and pressure, define
\begin{equation}
Q=\frac{x^2+y^2+4z^2}{2},
\qquad
H=Q+\frac{|\B|^2}{2}.
\label{eq:QH}
\end{equation}
A convenient flux label and the pressure are then
\begin{equation}
\psi=\frac{H-1+\eps^2/2}{2}
=\frac{x^2+y^2+4z^2+|\B|^2-2+\eps^2}{4},
\qquad
p=p_a-2\psi,
\label{eq:pressure}
\end{equation}
where $p_a$ is arbitrary.
(The normalization and offset of $\psi$ are chosen so it will vanish on the magnetic axis, and it will reduce to the poloidal flux divided by $2\pi$ in axisymmetry.)
The field is divergence-free and satisfies $\B\cdot\nabla\B=-\nabla Q$, as follows from the deformation described below.
The vector identity
\begin{equation}
(\nabla\times\B)\times\B
=\B\cdot\nabla\B-\nabla\frac{|\B|^2}{2}
=-\nabla H=\nabla p
\label{eq:force}
\end{equation}
then establishes exact scalar-pressure force balance.
In particular, $\B\cdot\nabla\psi=0$ follows from (\ref{eq:force}).
Equivalently, $\B$ is a steady incompressible Euler velocity with hydrodynamic pressure $Q$, and $H$ is its Bernoulli function.
The flux surfaces for these equilibria are shown in figure \ref{fig:geometry} for several choices of $\epsilon$.

The 3D solutions in this family can be understood as a deformation of an axisymmetric field.
Indeed, they continuously connect to an axisymmetric configuration as $\eps\to0$.
Then $a,b\to1$ and $s\to R^2$, so (\ref{eq:B_cartesian}) reduces, in cylindrical coordinates $(R,\phi,Z)$, to
\begin{equation}
\B_0=\frac{2Z}{R}\eR+\frac{I(\Psi_0)}{R}\ephi+(1-R^2)\eZ,
\qquad
\Psi_0=\frac{(R^2-1)^2}{4}+Z^2,
\label{eq:seed}
\end{equation}
where $I(\Psi_0)=\sqrt{1-4\Psi_0}$ and $\Psi_0<1/4$.
In this limit, $\psi\to\Psi_0$ and $p\to p_0=p_a-2\Psi_0$.
The polynomial $\Psi_0$ is a special case of equation (2.15) of \cite{solovev1968}.
It satisfies
\begin{equation}
\Delta^*\Psi_0=2R^2+2,
\qquad
p_0'=-2,
\qquad
II'=-2,
\label{eq:GS}
\end{equation}
where $\Delta^*=\partial_R^2-R^{-1}\partial_R+\partial_Z^2$ and primes denote derivatives with respect to $\Psi_0$.
Thus it solves the Grad--Shafranov equation with $p_0$ linear in $\Psi_0$.

The full three-dimensional solution can be obtained by stretching this axisymmetric field.
The key property of $\B_0$ that enables this transformation is
\begin{equation}
(\B_0\cdot\nabla)\B_0=-D\xvec,
\qquad
D=\operatorname{diag}(1,1,4),
\label{eq:accel_seed}
\end{equation}
where $\xvec=[x,y,z]^T$ is the Cartesian position.
Equivalently, the radial and vertical components of the left-hand side are $-R$ and $-4Z$.
With $a$ and $b$ as defined above, (\ref{eq:B_cartesian}) is precisely
\begin{equation}
\B(\xvec)=A\B_0(A^{-1}\xvec),
\qquad
A=\operatorname{diag}(a,b,1).
\label{eq:affine}
\end{equation}
Since $A$ is constant, the divergence remains zero.
Also, the chain rule gives $\B\cdot\nabla\B = A [\B_0\cdot\nabla\B_0](A^{-1}\xvec) =-ADA^{-1}\xvec=-D\xvec=-\nabla Q$, because $A$ commutes with $D$.


\subsection{Field line coordinates}
\label{sec:surfaces}

To understand the field lines in these 3D solutions, parameterize the position along a field line $\rvec(\zeta)$ using $\dot{\rvec} = \B(\rvec(\zeta))$, where dots denote $d/d\zeta$.
Since $\ddot{\rvec}=\B\cdot\nabla\B=-D\rvec$, each Cartesian coordinate satisfies a harmonic oscillator equation
\begin{equation}
  \ddot{x}=-x, \qquad
  \ddot{y}=-y, \qquad
  \ddot{z}=-4z.
  \label{eq:oscillators}
\end{equation}
Evidently then $\zeta$ has period $2\pi$, so it acts like a toroidal angle, though it differs generally from the standard angle.
The Cartesian coordinates therefore oscillate at frequencies 1, 1, and 2.
Requiring these oscillator solutions also to satisfy  $\dot{\rvec}=\B$ with (\ref{eq:B_cartesian})-(\ref{eq:auxiliary}), the solution can be written in terms of two constants $u$ and $v$ as follows.
The position map $\rvec(u,v,\zeta)$ is
\begin{equation}
\begin{aligned}
x&=a\left[L\cos \zeta+\frac{u\cos \zeta+v\sin \zeta}{L}\right],\\
y&=b\left[L\sin \zeta+\frac{v\cos \zeta-u\sin \zeta}{L}\right],\\
z&=v\cos 2\zeta-u\sin 2\zeta,
\end{aligned}
\label{eq:embedding}
\end{equation}
where
\begin{equation}
q^2=u^2+v^2,
\qquad
L=\left(\frac{1+\sqrt{1-4q^2}}{2}\right)^{1/2},
\qquad
L^2+\frac{q^2}{L^2}=1.
\label{eq:L}
\end{equation}
Direct substitution confirms $\B(\rvec)=\partial_\zeta\rvec$.
The parameters $(u,v)$ act as field line labels.
Differentiation gives
\begin{equation}
\det\frac{\partial(x,y,z)}{\partial(u,v,\zeta)}=-ab,
\label{eq:Jacobian}
\end{equation}
so the map is locally invertible.

It can be shown that for $u^2+v^2<1/4$, the coordinates in (\ref{eq:embedding}) cover $\mathcal{U}_\epsilon$ exactly once for a $2\pi$-period of $\zeta$.
Each position has a unique pair of field-line labels $(u,v)$, which vary smoothly with position.



\subsection{Pressure surfaces and magnetic axis}

The effective conserved ``energy'' equations for the harmonic oscillators (\ref{eq:oscillators}) can be obtained from (\ref{eq:embedding}) and its $d/d\zeta$ derivatives, with (\ref{eq:L}):
\begin{equation}
x^2+B_x^2=a^2(1+2u),
\quad
y^2+B_y^2=b^2(1-2u),
\quad
4z^2+B_z^2=4(u^2+v^2).
\label{eq:energies}
\end{equation}
Using $a^2-b^2=2\eps$ therefore yields
\begin{equation}
H=1+2\eps u+2(u^2+v^2),
\qquad
\psi=\left(u+\frac{\eps}{2}\right)^2+v^2.
\label{eq:psi_labels}
\end{equation}
In this form it is clear that $\psi$ is non-negative.
The gradient $\nabla\psi$ vanishes only on the field line $u=-\eps/2$, $v=0$, which is the magnetic axis at which $\psi=0$.
The flux surfaces have the explicit parameterization
\begin{equation}
\rvec\left(-\frac{\eps}{2}+\sqrt \psi\cos\alpha,\,
                 \sqrt \psi\sin\alpha,\,\zeta\right),
\qquad (\alpha,\zeta)\in S^1\times S^1.
\label{eq:surface}
\end{equation}


Now consider the range of allowed $\psi$.
The original domain $\mathcal{U}_{\epsilon}$ where the fields are analytic corresponds to $q<1/2$, which can be seen by squaring the $x$ and $y$ components in (\ref{eq:embedding}) to show $q^2=z^2+(s-1)^2/4$, and comparing to the definition of $\mathcal{U}_{\epsilon}$.
So, that domain corresponds to a disk of radius $1/2$ centered at the origin of the $u-v$ plane.
The $\psi$-surfaces are circles of radius $\sqrt\psi$ centered at $(-\epsilon/2,0)$ in this plane.
So, the $\psi$ surfaces remain within $\mathcal{U}_{\epsilon}$ if we choose an outermost surface $\psi=\delta$ where $\delta$ satisfies
\begin{equation}
0 < \delta<\frac{(1-\eps)^2}{4}.
\end{equation}
We thus arrive at the new domain
\begin{equation}
\Omega_\delta=\{\xvec\in\mathcal U_\eps:0\leq\psi\leq\delta\},
\label{eq:domain}
\end{equation}
which is a solid torus.
In contrast to $\mathcal{U}_{\epsilon}$, the domain $\Omega_{\delta}$ is bounded by a $\psi$ surface.

On the magnetic axis, (\ref{eq:embedding}) simplifies to
\begin{equation}
\boldsymbol\gamma(\zeta)=\left(
\sqrt{1-\eps^2}\cos \zeta,\,
\sqrt{1-\eps^2}\sin \zeta,\,
\frac{\eps}{2}\sin 2\zeta
\right).
\label{eq:axis}
\end{equation}
Hence the axis major radius $R_a=\sqrt{1-\eps^2}$ is constant, while the axis elevation $Z_a=(\epsilon/2)\sin 2\zeta$ is oscillatory.
For $\eps\ne0$, the axis is nonplanar.

The volume and toroidal flux of a pressure surface follow from (\ref{eq:Jacobian}) and (\ref{eq:psi_labels}).
For the region enclosed by a given value of $\psi$, corresponding to a cylinder of radius $\sqrt{\psi}$ and length $2\pi$ in $(u,v,\zeta)$ coordinates, the volume is
\begin{equation}
V(\psi)=2\pi^2a b\psi.
\label{eq:volume}
\end{equation}
The toroidal flux $\Phi_t$ can be computed as an integral over a constant-$\zeta$ surface:
\begin{equation}
\Phi_t(\psi)=
\int d\mathbf{A}\cdot\B
=\int du\int dv \left( -\partial_u \rvec \times \partial_v\rvec\right)\cdot\partial_\zeta\rvec =
\pi a b\psi,
\label{eq:flux}
\end{equation}
where the orientation of the normal was chosen to get a positive flux.


\subsection{Rotational transform}
\label{sec:transform}

We now compute the rotational transform for this family of solutions.
Since the $x$ and $y$ components of (\ref{eq:embedding}) are linear combinations of $\sin \zeta$ and $\cos\zeta$, a field line traces an ellipse around the origin in the $x-y$ plane once per period.
Together with the fact that the toroidal field is everywhere positive,
\begin{equation}
\B\cdot\nabla\phi=\frac{abF}{x^2+y^2}>0,
\label{eq:toroidal_rate}
\end{equation}
this implies that a $2\pi$ increase in $\zeta$ corresponds to one positive toroidal circuit.

Rotational transform can be measured using the geometric poloidal angle relative to the magnetic axis, which we define by 
\begin{equation}
\thetageo=\arg\left[(R-R_a)-i\{z-Z_a(\phi)\}\right].
\label{eq:poloidal_angle}
\end{equation}
First consider the axisymmetric limit $\eps=0$.
From (\ref{eq:embedding}) we find $R^2=x^2+y^2=1+2u\cos2\zeta+2v\sin2\zeta$.
Writing $u+iv=qe^{i\mu}$, we obtain
\begin{equation}
\frac{R^2-1}{2}-iz=qe^{i(2\zeta-\mu)}.
\label{eq:winding_seed}
\end{equation}
This displacement winds twice around the origin of the complex plane during one $\zeta$ period.
Replacing $(R^2-1)/2$ with $R-1$ does not change the number of turns around the origin, since the two differ by the positive factor $(R+1)/2$.
Since the axis has $R_a=1$ and $Z_a=0$, then (\ref{eq:poloidal_angle}) increases by $4\pi$  during a toroidal transit, so $\iota=2$.

For $\eps>0$, follow a field line with fixed $\psi>0$ and $\alpha$ in (\ref{eq:surface}) continuously from $\eps=0$ to its desired value.
The chosen field line stays within the domain (\ref{eq:domain}) along this deformation.
The field line always closes after one toroidal turn, since the position vector (\ref{eq:embedding}) is periodic in $\zeta$, so $\iota$ must remain an integer.
That integer could only change along the deformation if the field line crossed the magnetic axis, but this is impossible because the chosen value of $\psi$ for the field line remains fixed and different from the axis value $\psi=0$.
Consequently, on every surface in $\Omega_\delta$,
\begin{equation}
\Delta\phi=2\pi,
\qquad
\Delta\thetageo=4\pi,
\qquad
\iota=\frac{\Delta\thetageo}{\Delta\phi}=2.
\label{eq:iota}
\end{equation}


\subsection{Volume-averaged beta}
\label{sec:beta}

Next, we can derive explicit formulas for the averaged field strength, pressure, and $\beta$ (the ratio of thermal to magnetic pressure).
As a preliminary step, apply $\partial/\partial \zeta$ to (\ref{eq:embedding}), square, average over $\zeta$ at fixed $(u,v)$, and simplify using (\ref{eq:L}). The result can be written using (\ref{eq:psi_labels}) as
\begin{equation}
\frac{1}{2\pi}\int_0^{2\pi}|\B(\rvec(u,v,\zeta))|^2\,d\zeta
=1-\frac{\eps^2}{2}+2\psi.
\label{eq:period_average_B2}
\end{equation}
Now define the volume average over $\Omega_\delta$ as $\langle f\rangle_V=[V(\delta)]^{-1}\int_{\Omega_\delta}f\,d^3x$.
Combining (\ref{eq:period_average_B2}) with $\langle\psi\rangle_V=\delta/2$ (which follows from $\psi\propto V$ in  (\ref{eq:volume})), using the constant Jacobian (\ref{eq:Jacobian}) and the pressure (\ref{eq:pressure}), we obtain
\begin{equation}
\langle p\rangle_V=p_a-\delta,
\qquad
\langle|\B|^2\rangle_V=1-\frac{\eps^2}{2}+\delta.
\label{eq:volume_averages}
\end{equation}
The square root of this latter quantity is the usual definition of average field strength used in the stellarator literature, corresponding to \texttt{volavgB} in the widely used \texttt{VMEC} equilibrium code.
Writing $p_b=p(\delta)=p_a-2\delta$ for the boundary pressure, the ratio of volume-averaged thermal to magnetic pressure in the normalization of (\ref{eq:mhd}) is therefore
\begin{equation}
\beta_V\equiv\frac{2\langle p\rangle_V}{\langle|\B|^2\rangle_V}
=\frac{2(p_b+\delta)}{1-\eps^2/2+\delta}.
\label{eq:beta_volume_general}
\end{equation}
This quantity is the average $\beta$ commonly used in the stellarator literature, called \texttt{betatotal} in \texttt{VMEC}.
For pressure vanishing at the boundary, $p_a=2\delta$ and $p=2(\delta-\psi)$, so
\begin{equation}
\langle p\rangle_V=\delta,
\qquad
\beta_V=\frac{2\delta}{1-\eps^2/2+\delta}.
\label{eq:beta_volume_zero_edge}
\end{equation}
For the example in the first panel of figure \ref{fig:geometry}, this gives $\beta_V=2/57\simeq3.51\%$.


\section{Solutions with magnetic shear}
\label{sec:sheared_family}

We now consider a distinct family of solutions for which the rotational transform varies between flux surfaces.

\subsection{Magnetic field and pressure}
\label{sec:sheared_field}

Choose $\eps,S,\lambda>0$ and define the complex quantities
\begin{equation}
\omega=x+iy,
\qquad
K=\bar\omega\sqrt{1+\frac{\eps}{\bar\omega^2}},
\qquad
\Xi=\omega K+\frac{\pi}{2}-S.
\label{eq:sheared_auxiliary}
\end{equation}
The bar denotes complex conjugation, and the square root is chosen to have positive real part.
The field, flux label, and pressure are
\begin{equation}
\begin{aligned}
B_x&=\operatorname{Re}\left(\frac{ie^{-i\lambda z}}{2K}\sin\Xi\right),
\qquad
B_y=\operatorname{Im}\left(\frac{ie^{-i\lambda z}}{2K}\sin\Xi\right),\\
B_z&=\frac{1}{\lambda}\operatorname{Re}\left(e^{-i\lambda z}\cos\Xi\right),
\end{aligned}
\label{eq:sheared_B}
\end{equation}
\begin{equation}
\psi=\frac12\left[\sin^2(\lambda z)
+\left\{\operatorname{Re}\left(e^{-i\lambda z}\cos\Xi\right)\right\}^2\right],
\qquad
p=p_a-\frac{1}{\lambda^2}\psi,
\label{eq:sheared_pressure}
\end{equation}
where $p_a$ is arbitrary.
The parameter $\lambda$ sets the (inverse) vertical size of the flux surfaces, and also varies the pressure.
The parameters $S$ and $\eps$ affect the shape of the magnetic axis and the rotational transform, as shown below.
There is no restriction $\eps<1$ in this family.

To verify the equilibrium equations, we first check $\nabla\cdot\B$ as follows.
We can change variables from $(x,y)$ to $(\omega,\bar\omega)$ in partial derivatives using
\begin{equation}
\partial_\omega=\frac12(\partial_x-i\partial_y),
\qquad
\partial_{\bar\omega}=\frac12(\partial_x+i\partial_y).
\label{eq:sheared_W}
\end{equation}
The horizontal field can be written $B_x+iB_y=e^{-i\lambda z}W$ where
$W = (i \sin\Xi) / (2K)$.
Applying $\partial_\omega$, we find that $\operatorname{Re}(2e^{-i\lambda z}\partial_\omega W)$ equals the horizontal divergence $\partial_xB_x+\partial_yB_y$.
Also, since $K^2=\bar\omega^2+\eps$, we have $\partial_\omega K=0$ and $\partial_{\bar\omega}K=\bar\omega/K$.
Differentiating $W$ then gives $2\partial_\omega W=i\cos\Xi$.
Hence, $\partial_zB_z$ cancels the horizontal divergence found earlier, leaving $\nabla\cdot\B=0$.

Next we demonstrate force balance.
To this end, we compute the magnetic tension using
\begin{equation}
\B\cdot\nabla=e^{-i\lambda z}W\partial_\omega
+e^{i\lambda z}\bar W\partial_{\bar\omega}+B_z\partial_z.
\label{eq:B_dot_grad}
\end{equation}
The horizontal components of the tension $T_{x,y}=\B\cdot\nabla B_{x,y}$ can be computed using $i\lambda B_z=e^{-i\lambda z}\partial_{\omega} W-e^{i\lambda z}\overline{\partial_\omega W}$ and $\overline{\partial_\omega W}=\partial_{\bar\omega}\bar{W}$ with (\ref{eq:B_dot_grad}) to find $T_x + i T_y = \B\cdot\nabla(e^{-i\lambda z} W)=(\partial_x + i \partial_y) (|W|^2/2)$.
To compute the vertical component of the tension, we first evaluate
\begin{equation}
\partial_\omega^2 W=-K^2W,
\qquad
\partial_\omega\partial_{\bar\omega}W=-|\omega|^2W,
\qquad
W\,\partial_\omega^2 W-(\partial_\omega W)^2=\frac{1}{4}.
\label{eq:sheared_identities}
\end{equation}
Then the vertical component of $(\B\cdot\nabla)\B$ is $\B\cdot\nabla B_z = (2/\lambda)\operatorname{Im}\{ \B\cdot\nabla (e^{-i\lambda z}\partial_{\omega}W) \}=-(1/\lambda)\sin(\lambda z)\cos(\lambda z)$.
Thus $(\B\cdot\nabla)\B=-\nabla Q$, with
\begin{equation}
Q=-\frac{|W|^2}{2}+\frac{1}{2\lambda^2}\sin^2(\lambda z),
\qquad
H=Q+\frac{|\B|^2}{2}=\frac{1}{\lambda^2}\psi,
\label{eq:sheared_QH}
\end{equation}
where the last equality follows from $B_x^2+B_y^2=|W|^2$.
The vector identity (\ref{eq:force}) then gives $(\nabla\times\B)\times\B=-\nabla H=\nabla p$, and hence $\B\cdot\nabla\psi=0$.

Smoothness of these functions breaks down where $K=0$, which occurs at $(x=0,y=\pm\sqrt\epsilon)$, and on the branch cut where the radicand in $K$ is real and negative, which occurs where $x=0$ and $|y| < \sqrt{\epsilon}$.
So, the fields are smooth provided $(x,y)$ avoids the segment $x=0$, $|y|\leq\sqrt\eps$, as in the toroidal domain constructed below.

As $\eps\to0$, $K\to\bar\omega$ and $\Xi\to R^2+\pi/2-S$.
Using $B_R+iB_\phi=e^{-i\phi}(B_x+iB_y)$ then shows that the cylindrical field components become independent of $\phi$.
Thus, $\eps\to 0$ corresponds to axisymmetry.

\begin{figure}
\centering
\hspace{0.3in}
\includegraphics[width=0.4\textwidth]
{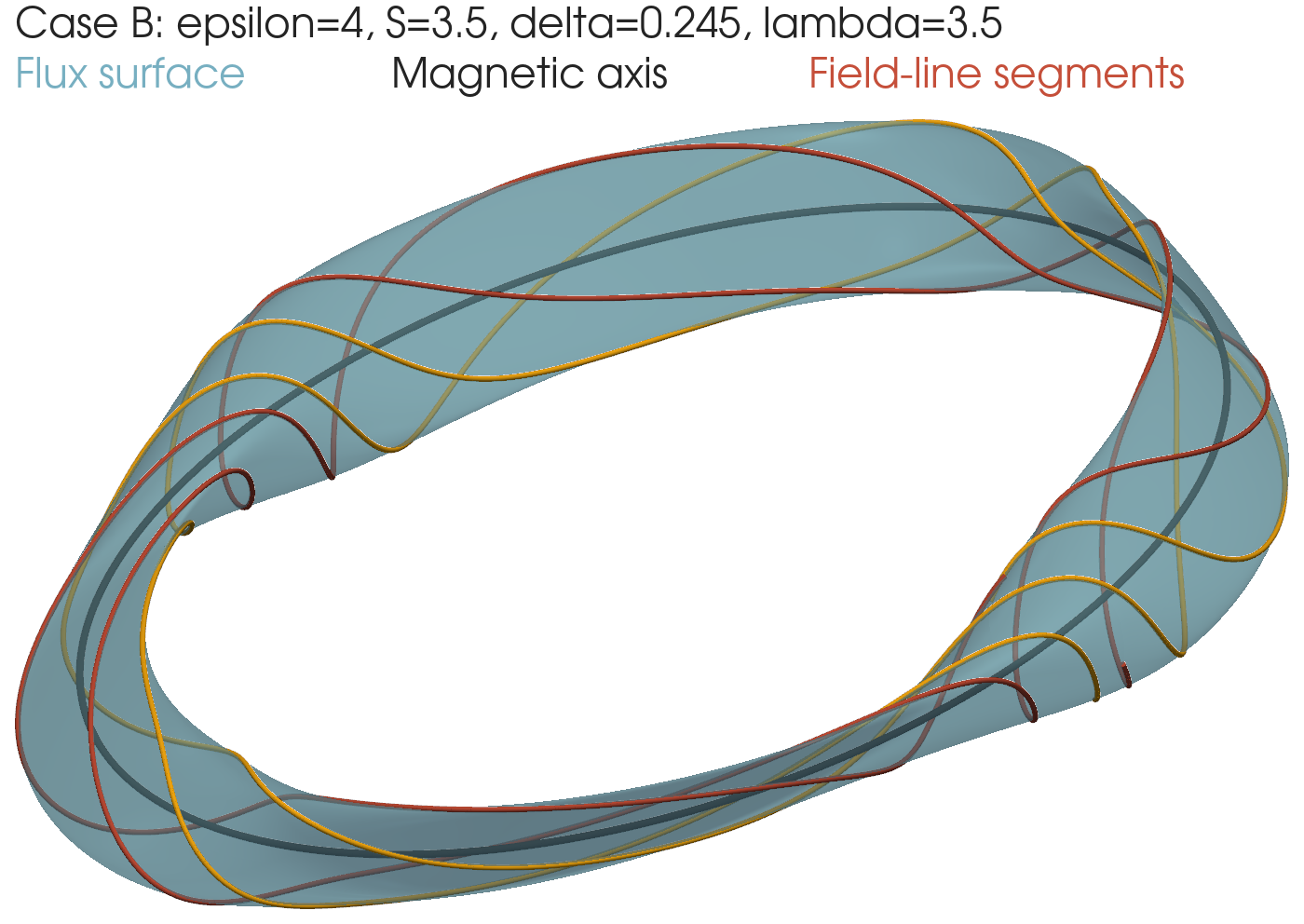}\hfill
\includegraphics[width=0.49\textwidth]{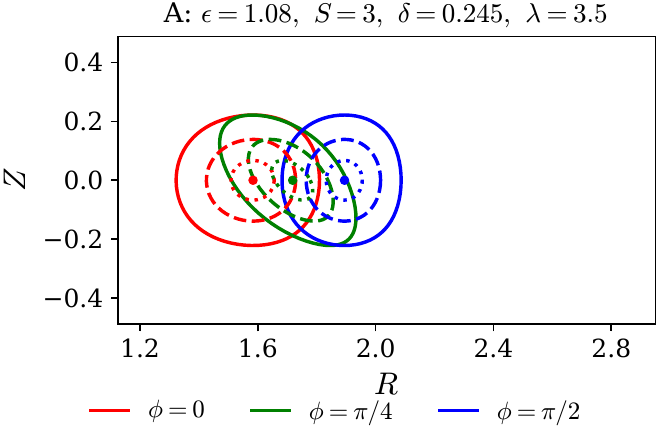}
\includegraphics[width=0.49\textwidth]{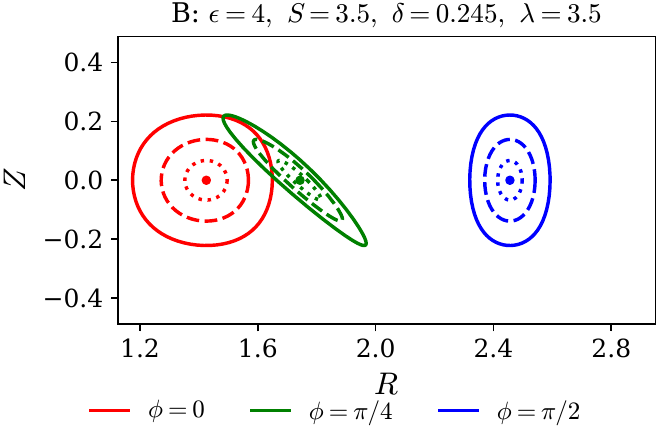}
\hfill
\includegraphics[width=0.49\textwidth]{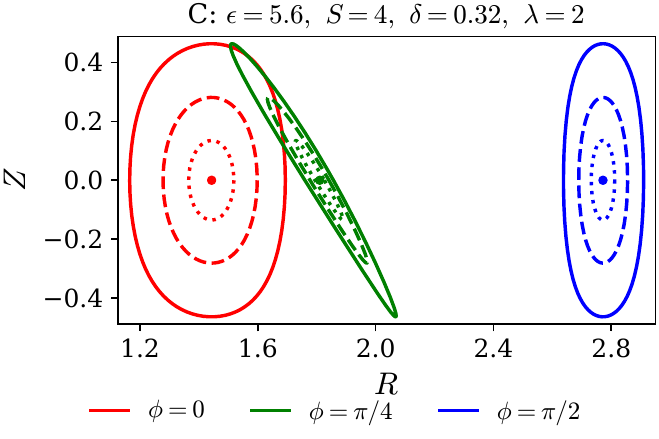}
\caption{The equilibrium family with sheared $\iota$.
Top left: 3D rendering.
Other panels: cross-sections of the flux surfaces for several choices of parameters.}
\label{fig:sheared}
\end{figure}

The fields in this family are stellarator-symmetric, with two field periods, as with the $\iota=2$ solutions.
The flux surface shapes for several choices of parameters are shown in figure \ref{fig:sheared}.

\subsection{Pressure surfaces and magnetic axis}
\label{sec:sheared_surfaces}

The form of (\ref{eq:sheared_pressure}) suggests using the coordinates
\begin{equation}
X=\lambda B_z,
\qquad
Y=-\sin(\lambda z),
\qquad
\psi=\frac{X^2+Y^2}{2}.
\label{eq:sheared_XY}
\end{equation}
The pressure surfaces are therefore circles in the $X$-$Y$ plane.
Unlike $u,v$ in section \ref{sec:surfaces}, $X,Y$ vary along a field line; only $X^2+Y^2$ is constant.

To convert these circles to surfaces in real space, introduce horizontal coordinates $\sigma,\zeta$ by $x=a_c(\sigma)\cos\zeta$, $y=b_c(\sigma)\sin\zeta$, so $\zeta$ is a toroidal angle, where
\begin{equation}
\begin{aligned}
a_c(\sigma)&=\sqrt{\frac{h(\sigma)-\eps}{2}},
&b_c(\sigma)&=\sqrt{\frac{h(\sigma)+\eps}{2}},
&h(\sigma)&=\sqrt{4\sigma^2+\eps^2}.
\end{aligned}
\label{eq:confocal_functions}
\end{equation}
For $\sigma>0$, the semiaxes satisfy $a_cb_c=\sigma$ and $b_c^2-a_c^2=\eps$, so constant-$\sigma$ curves are ellipses with foci at $(x,y)=(0,\pm\sqrt\eps)$.
Substituting these coordinates into (\ref{eq:sheared_auxiliary}) gives
\begin{equation}
\begin{alignedat}{3}
K&=b_c\cos\zeta-ia_c\sin\zeta,
&\qquad \omega K&=\sigma+i\nu,
&\qquad \nu(\zeta)&=\frac{\eps}{2}\sin2\zeta.
\end{alignedat}
\label{eq:sheared_confocal_identities}
\end{equation}
This formula for $K$ can be found by first computing $K^2$, then choosing the sign based on the real part of $K/\bar\omega$. 
Writing $\xi=\sigma-S$ and $\tau=-\lambda z$, and expanding the complex cosine in $B_z$, gives
\begin{equation}
X=-\sin\xi\cosh\nu\cos\tau+\cos\xi\sinh\nu\sin\tau,
\qquad Y=\sin\tau.
\label{eq:sheared_forward}
\end{equation}
Choose $|\tau|<\pi/2$, so $\cos\tau=\sqrt{1-Y^2}$.
The two terms in $X$ can be combined into a shifted sine of $\xi$, with amplitude $\sqrt{\cosh^2\nu-Y^2}$.
Solving the result for $\sigma$ gives
\begin{equation}
\sigma=S+\arctan\left(\frac{\tanh\nu\,Y}{\sqrt{1-Y^2}}\right)
-\arcsin\left(\frac{X}{\sqrt{\cosh^2\nu-Y^2}}\right).
\label{eq:sheared_sigma}
\end{equation}
Here $X^2+Y^2<1$, and the inverse trigonometric functions take values between $-\pi/2$ and $\pi/2$.
The position is therefore given explicitly by
\begin{equation}
\rvec(X,Y,\zeta)=\left(a_c(\sigma)\cos\zeta,\,
b_c(\sigma)\sin\zeta,\,-\frac{\arcsin Y}{\lambda}\right),
\label{eq:sheared_embedding}
\end{equation}
with $\sigma(X,Y,\nu(\zeta))$ given by (\ref{eq:sheared_sigma}).
Applying this map to circles of constant $X^2+Y^2$ for $\zeta\in[0,2\pi)$ then gives the explicit shapes of the flux surfaces.

We next examine the range of allowed $\psi$.
Squaring (\ref{eq:sheared_forward}) and rearranging gives
\begin{equation}
2\psi=\sin^2\xi+
\left(\cos\xi\cosh\nu\sin\tau-\sin\xi\sinh\nu\cos\tau\right)^2.
\label{eq:sheared_domain_bound}
\end{equation}
Hence $2\psi\geq\sin^2(\sigma-S)$.
For the surfaces surrounding $\sigma=S$, choosing an outermost surface $\psi=\delta$ with
\begin{equation}
k_b=\sqrt{2\delta}<1,
\qquad
\delta>0,
\qquad
\arcsin k_b < S,
\label{eq:sheared_domain}
\end{equation}
ensures $|\sigma-S|\leq\arcsin k_b<S$.
Thus $\sigma$ stays positive, keeping the surfaces away from the excluded region mentioned above ($x=0$, $|y| \le \sqrt\epsilon$), while $|\lambda z|\leq\arcsin k_b<\pi/2$.
The plasma domain $\Omega_\delta$ is obtained by allowing $X^2+Y^2\leq k_b^2$ and a full period of $\zeta$ in (\ref{eq:sheared_embedding}).

The coordinate Jacobian can be calculated by first defining
\begin{equation}
G(\sigma,\zeta)=|K|^2
=b_c^2 \cos^2\zeta + a_c^2 \sin^2\zeta = \frac{h(\sigma) + \epsilon \cos 2\zeta}{2},
\label{eq:G}
\end{equation}
and observing $\det\partial(x,y)/\partial(\sigma,\zeta)=G/h$.
Combining this result with 
\begin{equation}
\left(\frac{\partial\sigma}{\partial X}\right)_{Y,\zeta}=-\frac{1}{T},
\qquad
T=\sqrt{\cosh^2\nu-X^2-Y^2},
\label{eq:T}
\end{equation}
gives
\begin{equation}
\det\frac{\partial(x,y,z)}{\partial(X,Y,\zeta)}
=-\frac{G}{\lambda hT\sqrt{1-Y^2}}<0.
\label{eq:sheared_Jacobian}
\end{equation}
This Jacobian is finite and nonzero throughout the chosen domain, including at $X=Y=0$.
The coordinates cover $\Omega_\delta$ once, so circles in the $X$-$Y$ plane describe nested smooth tori, with smooth field, current, and pressure.
In particular, each surface $\psi=k^2/2$ can be plotted by substituting
\begin{equation}
X=-k\cos\chi,
\qquad
Y=k\sin\chi,
\qquad 0<k\leq k_b
\label{eq:sheared_chi}
\end{equation}
into (\ref{eq:sheared_embedding}) and varying both angles $\chi,\zeta$ over $2\pi$.

Since $\psi=(X^2+Y^2)/2$, $\nabla\psi$ and $\nabla p$ vanish only at $X=Y=0$.
There $\sigma=S$ and $z=0$, giving the magnetic axis
\begin{equation}
\boldsymbol\gamma(\zeta)=\left(a_c(S)\cos\zeta,\,b_c(S)\sin\zeta,\,0\right).
\label{eq:sheared_axis}
\end{equation}
The axis is a planar ellipse, reducing to a circle of radius $\sqrt S$ as $\eps\to0$.


\subsection{Rotational transform}
\label{sec:sheared_transform}

For rotational transform, use the poloidal angle $\chi$ in (\ref{eq:sheared_chi}).
At $\chi=0$, $z=0$ and $\sigma>S$, so the point lies outside the axis.
Increasing $\chi$ initially decreases $z$, matching the orientation of the geometric angle (\ref{eq:poloidal_angle}).

Differentiating $Y=-\sin(\lambda z)$ along $\B$ gives
$\B\cdot\nabla Y=-\lambda B_z\cos(\lambda z)=-X\sqrt{1-Y^2}$.
Substitute (\ref{eq:sheared_chi}), with $k$ constant along the field, to obtain 
\begin{equation}
\B\cdot\nabla\chi=\sqrt{1-k^2\sin^2\chi}.
\label{eq:B_dot_grad_chi}
\end{equation}
To get the analogous toroidal component, first differentiate the horizontal coordinate map to get
\begin{equation}
\nabla\zeta=\frac{(-a_c\sin\zeta,\,b_c\cos\zeta,\,0)}{G}.
\end{equation}
Taking its dot product with (\ref{eq:sheared_B}) and using (\ref{eq:sheared_confocal_identities}) gives $\B\cdot\nabla\zeta=G^{-1}\operatorname{Im}([B_x + i B_y]K) = (2G)^{-1} \operatorname{Re}(e^{i\tau}\sin\Xi)$.
Using $\Xi=\sigma+i\nu-S+\pi/2$, then $\B\cdot\nabla\zeta=-(2G)^{-1}(\partial X/\partial\sigma)_{z,\zeta}$.
The inverse of (\ref{eq:T}) then yields
\begin{equation}
\B\cdot\nabla\zeta=\frac{1}{2G}\sqrt{\cosh^2\nu-k^2}.
\label{eq:B_dot_grad_zeta}
\end{equation}
Divide (\ref{eq:B_dot_grad_chi}) by (\ref{eq:B_dot_grad_zeta}), and count poloidal turns over many toroidal transits:
\begin{equation}
\begin{aligned}
\frac{d\chi}{d\zeta}
&=\frac{2G(\sigma,\zeta)\sqrt{1-k^2\sin^2\chi}}
{\sqrt{\cosh^2\nu(\zeta)-k^2}},\\
\iota(\psi)&=\lim_{N\to\infty}\frac{\chi(\zeta+2\pi N)-\chi(\zeta)}{2\pi N}.
\end{aligned}
\label{eq:sheared_iota}
\end{equation}
Here $\sigma$ is evaluated from (\ref{eq:sheared_sigma}) and (\ref{eq:sheared_chi}).
Since $a_c,b_c>0$, $(x,y)$ passes through its four quadrants in order as $\zeta$ increases by $2\pi$.
Together with $\B\cdot\nabla\zeta>0$, this gives one positive toroidal transit.
The parameter $\lambda$ does not appear in (\ref{eq:sheared_iota}), so changing it leaves $\iota(\psi)$ unchanged.

On the magnetic axis $k\to0$, so $\sigma\to S$ and $d\chi/d\zeta=[h_0+\eps\cos2\zeta]\operatorname{sech}\nu$, where $h_0=h(S)$.
The cosine term averages to zero, since $\zeta\mapsto\pi/2-\zeta$ leaves $\nu$ unchanged but reverses $\cos2\zeta$.
Defining
\begin{equation}
\mathcal H(\eps)=\frac{1}{2\pi}\int_0^{2\pi}\operatorname{sech}\nu(\zeta)\,d\zeta,
\label{eq:sheared_shaping_average}
\end{equation}
we therefore obtain the on-axis transform
\begin{equation}
\iota(0)=h_0\mathcal H(\eps).
\label{eq:sheared_iota_axis}
\end{equation}

Away from the axis, the $k$-dependent factors in (\ref{eq:sheared_iota}) will generally make $\iota$ vary from surface to surface.
This variation enters through the explicit $k^2$ factors in (\ref{eq:sheared_iota}), as well as through the dependence of $G$ on $\sigma$ via (\ref{eq:G}) and (\ref{eq:sheared_sigma}), which depends on $k$ via (\ref{eq:sheared_chi}).
This dependence of $\iota$ on $k$ is smooth within the domain, and $\iota(k)$ is generally non-constant, so $\iota$ will take on both rational and irrational values.
Consequently, these fields have both closed and non-closed field lines.

Given specific parameters, (\ref{eq:sheared_iota}) can be evaluated numerically for each $k$ to obtain the transform profile.
For example, the parameters
\begin{equation}
\eps=2,
\qquad S=1,
\qquad k_b=0.1,
\qquad \delta=\frac{1}{200}
\label{eq:sheared_example}
\end{equation}
give $\iota(0)\simeq2.28690$ and $\iota(\delta)\simeq2.2878$.


\subsection{Volume-averaged beta}
\label{sec:sheared_beta}

Next, we compute the averaged pressure, field strength, and $\beta_V$, defined in section \ref{sec:beta}, since these quantities provide additional useful benchmarks for equilibrium codes.
Write $p_b=p(\delta)=p_a-\delta/\lambda^2$ for the boundary pressure and define
\begin{equation}
\mathcal M=\langle\psi\rangle_V,
\qquad
\mathcal C=\langle X^2\rangle_V,
\qquad
\mathcal A=\langle B_x^2+B_y^2\rangle_V.
\label{eq:sheared_beta_coefficients}
\end{equation}
These coefficients are independent of $\lambda$: $B_x^2+B_y^2=|W|^2$, and the common factor $1/\lambda$ in the volume element cancels in normalized averages.
Averaging (\ref{eq:sheared_pressure}) and using $B_z^2=X^2/\lambda^2$ gives
\begin{equation}
\langle p\rangle_V=p_b+\frac{1}{\lambda^2}(\delta-\mathcal M),
\qquad
\langle|\B|^2\rangle_V=\mathcal A+\frac{\mathcal C}{\lambda^2}.
\label{eq:sheared_volume_averages}
\end{equation}
Hence
\begin{equation}
\beta_V=\frac{2\langle p\rangle_V}{\langle|\B|^2\rangle_V}
=\frac{2\lambda^2p_b+2\delta-2\mathcal M}
{\lambda^2\mathcal A+\mathcal C}.
\label{eq:sheared_beta_general}
\end{equation}
For pressure vanishing at the boundary, $p=(\delta-\psi)/\lambda^2$, so
\begin{equation}
\beta_V=\frac{2(\delta-\mathcal M)}{\lambda^2\mathcal A+\mathcal C}.
\label{eq:sheared_beta_zero_edge}
\end{equation}
In this case $\beta_V$ decreases with $\lambda$ at fixed $\eps,S,\delta$.

These coefficients can be evaluated by first integrating analytically over height $\tau=-\lambda z$, leaving two numerical integrals.
Using $\xi=\sigma-S$ as before, define
\begin{equation}
\begin{aligned}
d&=\arcsin k_b,
&\mathcal F(\xi,\zeta)&=\cosh^2\nu(\zeta)-\sin^2\xi,\\
U(\xi,\zeta)&=\arcsin\sqrt{\frac{k_b^2-\sin^2\xi}{\mathcal F(\xi,\zeta)}},
&\tau_0(\xi,\zeta)&=\arctan(\tan\xi\tanh\nu).
\end{aligned}
\label{eq:sheared_beta_quadrature_aux}
\end{equation}
Combining the sine and cosine inside the square in (\ref{eq:sheared_domain_bound}) gives $2\psi=\sin^2\xi+\mathcal F\sin^2(\tau-\tau_0)$.
Thus at each horizontal position $(\xi,\zeta)$, $\tau$ ranges from $\tau_0-U$ to $\tau_0+U$, with $-d\leq\xi\leq d$.
The volume element is $d^3x=G\,d\xi\,d\zeta\,d\tau/(\lambda h)$, using the horizontal Jacobian $G/h$ derived above.
Also, (\ref{eq:sheared_B}), (\ref{eq:sheared_confocal_identities}), and (\ref{eq:G}) give $B_x^2+B_y^2=\mathcal F/(4G)$.

Define
\begin{equation}
\begin{aligned}
\mathcal I[f]&=\int_0^{2\pi}d\zeta\int_{-d}^{d}f(\xi,\zeta)\,d\xi,
&D_U&=U-\tfrac12\sin2U,\\
c_0&=-\frac{\sin\xi\cos\xi}{\sqrt{\mathcal F}},
&c_1&=\frac{\cosh\nu\sinh\nu}{\sqrt{\mathcal F}}.
\end{aligned}
\label{eq:sheared_beta_integrals}
\end{equation}
The height integrals follow by writing $X=c_0\cos(\tau-\tau_0)+c_1\sin(\tau-\tau_0)$, from (\ref{eq:sheared_forward}).
Over the interval $\tau_0-U\leq\tau\leq\tau_0+U$, the integrals of $\sin^2(\tau-\tau_0)$ and $\cos^2(\tau-\tau_0)$ are $D_U$ and $2U-D_U$, respectively, while their mixed product integrates to zero.
For the volume, pressure, and $X^2$ integrals, the term proportional to $\cos2\zeta$ in $G/h=1/2+\eps\cos2\zeta/(2h)$ cancels on averaging over $\zeta$, by the same symmetry used for (\ref{eq:sheared_iota_axis}).
For the horizontal magnetic energy, the factor $G$ instead cancels directly.
The resulting expressions are
\begin{equation}
\begin{aligned}
V(\delta)&=\frac{\mathcal I[U]}{\lambda},\\
\mathcal M&=\frac{\mathcal I[2\sin^2\xi\,U+\mathcal F D_U]}{4\mathcal I[U]},\\
\mathcal C&=\frac{\mathcal I[c_0^2(2U-D_U)+c_1^2D_U]}{2\mathcal I[U]},\\
\mathcal A&=\frac{\mathcal I[\mathcal F U/h(S+\xi)]}{2\mathcal I[U]}.
\end{aligned}
\label{eq:sheared_beta_quadratures}
\end{equation}


\section{Discussion and conclusions}
\label{sec:conclusions}

In this paper we have presented two families of explicit non-axisymmetric MHD equilibria in toroidal geometry with nested toroidal flux surfaces.
Both families have order-unity deviations from axisymmetry, with no expansion in the inverse aspect ratio or the distance from the magnetic axis.
Both families also have a non-vanishing pressure gradient everywhere except the magnetic axis.
The first family has a nonplanar magnetic axis and uniform rotational transform equal to an integer.
The second family has a sheared profile of rotational transform, with $\iota$ taking on both rational and irrational values.
In both families, the expressions for the magnetic field, flux surfaces, and pressure in Cartesian coordinates involve only elementary functions.
Together with the family in \cite{gomezserrano2026},
these solutions provide counterexamples to the formulation of Grad's conjecture in, for example, Conjecture 2.5 in \cite{cardona2025asymmetry}.
These analytic solutions thus advance our understanding of the existence and regularity of 3D equilibria.
For the stellarator fusion concept in particular, it is reassuring to know with certainty that strongly asymmetric equilibria with perfect flux surfaces do in principle exist.
We also expect the analytic solutions in this paper will be practically useful as tests of numerical equilibrium codes.
Demonstrations of tests of this type are provided in the supplemental materials.

Several extensions of this work are of interest for future research.
Since at least three families of exact solutions exist, it is likely that there are others.
Similar analytic solutions could be sought with other numbers of field periods, or with $\iota$ having a constant value that is irrational.
If explicit solutions could be found that break stellarator symmetry, they would be useful for testing the functionality for non-stellarator-symmetric fields in numerical codes.
It remains unknown whether exact analytic solutions exist for non-axisymmetric toroidal magnetic fields with nested toroidal flux surfaces if we demand the fields be curl-free (vacuum) or force-free.
Finally, a related open question is the existence of non-axisymmetric equilibria satisfying exact quasisymmetry.


\section*{Acknowledgements}
These solutions were discovered using the artificial intelligence model GPT-6 Astra Pro, which was also used to draft parts of the manuscript. All equations were confirmed manually by the author.

Thank you to Nathan Duignan and Daniel Ginsberg for feedback on the manuscript.

\section*{Funding}
This work was supported by the US Department of Energy under contract DE-FG02-93ER54197.

\section*{Declaration of interests}
The author is a consultant for Type One Energy Group.

\section*{Data availability statement}
Supplemental data associated with this work can be found at \citep{github}.

\bibliographystyle{jpp}
\bibliography{analytic_3d_equilibrium}

@string{PFB =   "Phys. Fluids B"}

@string{NF =   "Nucl. Fusion"}

@string{JETP =   "Soviet Phys. JETP"}

@article{grad1967,
  author = {Grad, H.},
  title = {Toroidal containment of a plasma},
  journal = {Phys. Fluids},
  volume = {10}, number = {1}, pages = {137--154}, year = {1967},
  doi = {10.1063/1.1761965}
}

@article{grad1985theory,
  title={Theory and applications of the nonexistence of simple toroidal plasma equilibrium},
  author={Grad, Harold},
  year={1985},
  journal={Int. J. Fusion Energy},
  volume=3,
  pages={33-46}
}

@article{solovev1968,
  author = {Solov'ev, L. S.},
  title = {The theory of hydromagnetic stability of toroidal plasma configurations},
  journal = {Sov. Phys. JETP},
  volume = {26}, number = {2}, pages = {400--407}, year = {1968},
  url = {https://www.jetp.ras.ru/cgi-bin/dn/e_026_02_0400.pdf}
}

@article{lortz1970,
  author = {Lortz, D.},
  title = {{\"U}ber die {Existenz} toroidaler magnetohydrostatischer {Gleichgewichte} ohne {Rotationstransformation}},
  journal = {Z. Angew. Math. Phys.},
  volume = {21}, pages = {196--211}, year = {1970},
  doi = {10.1007/BF01590644}
}

@article{kaiser1997,
  author = {Kaiser, R. and Salat, A.},
  title = {New classes of three-dimensional ideal-{MHD} equilibria},
  journal = {J. Plasma Phys.},
  volume = {57}, number = {2}, pages = {425--448}, year = {1997},
  doi = {10.1017/S0022377896004862}
}

@article{bruno1996,
  author = {Bruno, O. P. and Laurence, P.},
  title = {Existence of three-dimensional toroidal {MHD} equilibria with nonconstant pressure},
  journal = {Commun. Pure Appl. Math.},
  volume = {49}, number = {7}, pages = {717--764}, year = {1996},
  doi = {10.1002/(SICI)1097-0312(199607)49:7<717::AID-CPA3>3.0.CO;2-C},
  url = {https://authors.library.caltech.edu/records/hv51p-nyg34}
}

@article{weitzner2020,
  author = {Weitzner, H. and Sengupta, W.},
  title = {Exact non-symmetric closed line vacuum magnetic fields in a topological torus},
  journal = {Phys. Plasmas},
  volume = {27}, number = {2}, pages = {022509}, year = {2020},
  doi = {10.1063/1.5126688},
  url = {https://arxiv.org/abs/1909.01890}
}

@article{plunk2020perturbing,
  title={Perturbing an axisymmetric magnetic equilibrium to obtain a quasi-axisymmetric stellarator},
  author={Plunk, GG},
  journal={Journal of Plasma Physics},
  volume={86},
  number={4},
  pages={905860409},
  year={2020},
  publisher={Cambridge University Press}
}

@article{enciso2025,
  author = {Enciso, A. and Luque, A. and Peralta-Salas, D.},
  title = {{MHD} equilibria with nonconstant pressure in nondegenerate toroidal domains},
  journal = {J. Eur. Math. Soc.},
  volume = {27}, number = {6}, pages = {2251--2291}, year = {2025},
  doi = {10.4171/JEMS/1410}
}

@article{sorokina_ilgisonis2024,
  author = {Sorokina, E. A. and Ilgisonis, V. I.},
  title = {Existence of true plasma equilibria in asymmetric magnetic fields},
  journal = {Phys. Rev. E},
  volume = {110}, number = {6}, pages = {065209}, year = {2024},
  doi = {10.1103/PhysRevE.110.065209}
}

@misc{drivas2025,
  author = {Drivas, T. D. and Elgindi, T. M. and Ginsberg, D.},
  title = {On the existence of fibered three-dimensional perfect fluid equilibria without continuous {Euclidean} symmetry},
  year = {2025},
  howpublished = {Preprint},
  eprint = {2510.02955},
  doi = {10.48550/arXiv.2510.02955},
  url = {https://arxiv.org/abs/2510.02955}
}

@article{helander2014theory,
  title={Theory of plasma confinement in non-axisymmetric magnetic fields},
  author={Helander, Per},
  journal={Reports on Progress in Physics},
  volume={77},
  number={8},
  pages={087001},
  year={2014},
  publisher={IOP Publishing}
}

@article{KruskalKulsrud,
    author = {Kruskal, M. D. and Kulsrud, R. M.},
    title = {Equilibrium of a Magnetically Confined Plasma in a Toroid},
    journal = {The Physics of Fluids},
    volume = {1},
    number = {4},
    pages = {265-274},
    year = {1958},
    month = {07},
    doi = {10.1063/1.1705884},
    url = {https://doi.org/10.1063/1.1705884},
}

@article{cardona2025asymmetry,
  title={Asymmetry of {MHD} Equilibria for Generic Adapted Metrics},
  author={Cardona, Robert and Duignan, Nathan and Perrella, David},
  journal={Archive for Rational Mechanics and Analysis},
  volume={249},
  number={1},
  pages={1},
  year={2025},
  publisher={Springer}
}

@Article{Mercier,
  author =       {C Mercier},
  title="Equilibrium and stability of a toroidal magnetohydrodynamic system in the neighbourhood of a magnetic axis",
  journal =      NF,
  year =         {1964},
  volume =       {4},
  pages =        {213},
}

@Article{GB1,
  author =       {D A Garren and A H Boozer},
  journal =      PFB,
  year =         {1991},
  volume =       {3},
  pages =        {2805},
  title="Magnetic field strength of toroidal plasma equilibria"
}

@Book{SolovevShafranov,
author={L S Solov'ev and V D Shafranov},
address = {New York - London},
publisher = {Consultants Bureau},
title = {{Reviews of Plasma Physics 5}},
year = {1970}
}

@article{constantin2021flexibility,
  title={Flexibility and rigidity in steady fluid motion},
  author={Constantin, Peter and Drivas, Theodore D and Ginsberg, Daniel},
  journal={Communications in Mathematical Physics},
  volume={385},
  number={1},
  pages={521--563},
  year={2021},
  publisher={Springer}
}

@article{peralta2026symmetry,
  title={A symmetry theorem for localizable steady solutions of the {3D Euler} equations},
  author={Peralta-Salas, Daniel and Slobodeanu, Radu},
  journal={arXiv preprint arXiv:2606.13462},
  year={2026}
}

@article{Moffatt_1985,
author={Moffatt, H. K.}, 
title={Magnetostatic equilibria and analogous {Euler} flows of arbitrarily complex topology. {Part} 1. {Fundamentals}},
volume={159},
DOI={10.1017/S0022112085003251},
journal={Journal of Fluid Mechanics},
year={1985}, 
pages={359–378}
}

@article{constantin2023magnetic,
  title={Magnetic relaxation of a Voigt--MHD system},
  author={Constantin, Peter and Pasqualotto, Federico},
  journal={Communications in Mathematical Physics},
  volume={402},
  number={2},
  pages={1931--1952},
  year={2023},
  publisher={Springer}
}

@article{gomezserrano2026,
  title={Counterexamples to Grad's conjecture},
  author={Javier Gómez-Serrano and Lukas Liehr and Mitchell A Taylor},
  journal={arXiv preprint arXiv:2609.24739},
  year={2026}
}

@misc{github,
  author = {Landreman, M.},
  title = {Analytic 3D equilibria},
  howpublished = {\url{https://github.com/landreman/analytic_3d_equilibria}},
  year = {2026}
}
\end{document}